\documentclass[Physsubmission, Phys]{SciPost}

\hypersetup{
    colorlinks,
    linkcolor={red!50!black},
    citecolor={blue!50!black},
    urlcolor={blue!80!black}
}

\usepackage[bitstream-charter]{mathdesign}
\DeclareSymbolFont{usualmathcal}{OMS}{cmsy}{m}{n}
\DeclareSymbolFontAlphabet{\mathcal}{usualmathcal}

\begin{document}

\begin{center}{\Large \textbf{
Measurements of $\phi$ meson and $\Xi^-$ hyperon production in Au+Au collisions at $\sqrt{s_{NN}}$ = 3 GeV from STAR experiment\\
}}\end{center}

 \begin{center}
Yingjie Zhou\textsuperscript{1$\star$} (for the STAR Collaboration)
 \end{center}

\begin{center}
{\bf 1} Key Laboratory of Quark and Lepton Physics (MOE) and Institution of Particle Physics, Central China Normal University, Wuhan 430079, P. R. China
\\
* yingjiezhou@mails.ccnu.edu.cn
\end{center}

\begin{center}
\today
\end{center}


\definecolor{palegray}{gray}{0.95}
\begin{center}
\colorbox{palegray}{
  \begin{tabular}{rr}
  \begin{minipage}{0.1\textwidth}
    \includegraphics[width=30mm]{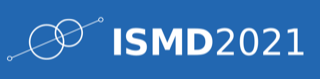}
  \end{minipage}
  &
  \begin{minipage}{0.75\textwidth}
    \begin{center}
    {\it 50th International Symposium on Multiparticle Dynamics}\\ {\it (ISMD2021)}\\
    {\it 12-16 July 2021} \\
    \doi{10.21468/SciPostPhysProc.?}\\
    \end{center}
  \end{minipage}
\end{tabular}
}
\end{center}

\section*{Abstract}
{\bf
In this proceedings, we present our first measurements of $\phi$ meson and $\Xi^-$ hyperon production in Au+Au collisions at $\sqrt{s_{NN}}$ = 3 GeV. Various models including thermal and transport model calculations are compared to data, these results imply that the matter produced in the 3 GeV Au+Au collisions is considerably different from that at higher energies.
}

\vspace{10pt}
\noindent\rule{\textwidth}{1pt}
\tableofcontents\thispagestyle{fancy}
\noindent\rule{\textwidth}{1pt}
\vspace{10pt}

\section{Introduction}
\label{sec:intro}
The main goal of the STAR experiment is to study the properties of QCD matter at extreme conditions, high temperature and/or high density, by colliding heavy ions at ultra-relativistic speed. The yields and particle ratios of strange hadrons provide important information about the particle production mechanisms in these collisions. The RHIC Beam Energy Scan program covers a wide range of energies to explore the transition from a hadronic dominated phase to a partonic dominated one. Of particular interest is the high baryon density region which is accessible through the STAR fixed-target program, which has extended the energy reach from $\sqrt{s_{NN}}$ = 13.7 GeV down to 3.0 GeV. 

Statistical thermal models have often been used to characterize the thermal properties of the produced media. In low energies, the strangeness production is rare, therefore it has been argued that strangeness number needs to be conserved locally on an event-by-event basis described by the Canonical Ensemble (CE), which leads to a reduction in the yields of hadrons with non-zero strangeness number \cite{Rafelski_1980279, Rafelski_2002,Redlich:2001kb}, but not for the $\phi$ meson with zero net strangeness number (S=0). The $\phi/K^-$ ratio is expected to increase with decreasing collision energy in models using the CE treatment for strangeness, opposite to the trend in the Grand Canonical Ensemble (GCE) treatment. The canonical suppression power for $\Xi^-$ (S=2) is even larger than for $K^-$ (S=1). The $\phi/K^-$ and $\phi/\Xi^-$ ratios offer a unique test to scrutinize thermodynamic properties of strange quarks in the hot and/or dense QCD environment \cite{STAR:2021hyx}. 

In this proceedings, the first measurements of $\phi$ meson and $\Xi^-$ hyperon production as well as the ratios of $\phi/K^-$ and $\phi/\Xi^-$ in Au+Au collisions at $\sqrt{s_{NN}}$ = 3 GeV will be presented. 

\section{Experiment}

The dataset used in this analysis consists of Au+Au collisions at $\sqrt{s_{NN}}$ = 3 GeV collected by the STAR experiment under the fixed target (FXT) configuration in the year of 2018. The single beam was provided by RHIC with total energy equal to 3.85 GeV/nucleon. The thickness of the gold target is about 0.25 mm, corresponding to a 1\% interaction probability to minimize the pileup and energy loss effect in target. The target is located at 200 cm to the west of the center of the STAR detector, it is installed inside the vacuum pipe, 2 cm below the center of the beam axis. The minimum bias (MB) trigger condition is provided by the Beam-Beam Counters (BBC). Tracking and particle identification (PID) are done using the energy loss (dE/dx) information from Time Projection Chamber (TPC) and time of flight ($1/\beta$) information from Time of Flight (TOF). In total, approximately $2.6\times10^8$ MB triggered events are used in this analysis.

\section{Particle yields}
\label{sec:particle}
 \begin{figure}[h!]
 \centering
 \includegraphics[width=0.65\textwidth]{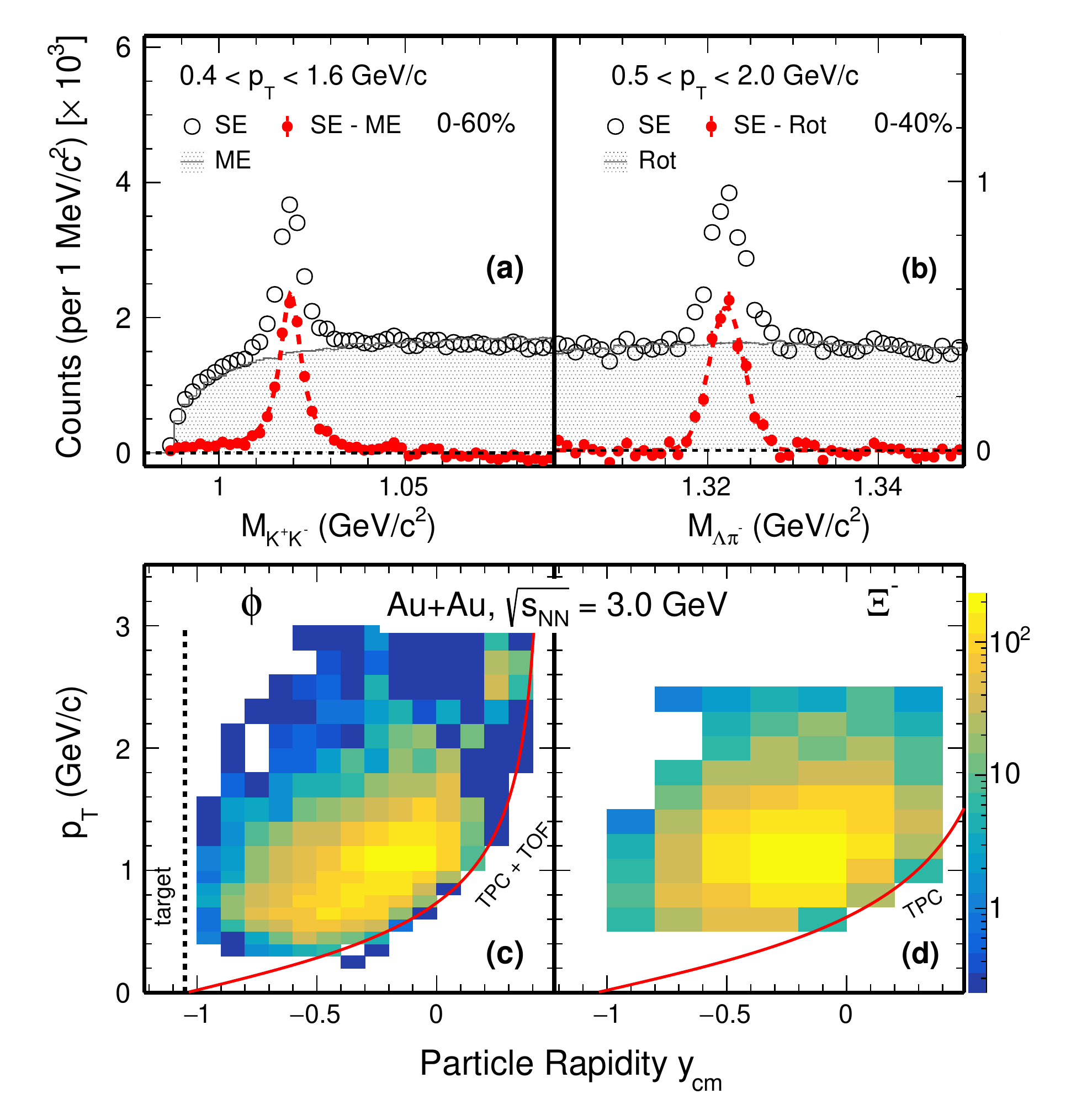}
 \caption{Invariant mass distributions of (a) $K^{+}K^{-}$ and (b) $\Lambda\pi^-$ pairs reconstructed from data are shown on the top. The grey shaded histogram represents the normalized mixed-event (rotating daughters for $\Xi^-$) unlike-sign distribution that is used to estimate the combinatorial background. The transverse momentum ($p_{\rm T}$) versus the rapidity ($y_{\rm cm}$) for reconstructed (c) $\phi$ and (d) $\Xi^-$ are shown on the bottom.}
 \label{fig:acc}
 \end{figure}

Reconstruction of short lived particles $\Xi^-\rightarrow\Lambda+\pi^-$ is performed using the KF Particle Finder package based on the Kalman Filter method \cite{Kisel:2018nvd}. Those combinatorial backgrounds are obtained by rotating daughter tracks. The $\phi$ mesons are reconstructed through the hadronic decay channel, $\phi\rightarrow K^{+} + K^-$, where combinatorial background is estimated using the mixed event technique. The reconstructed $\Xi^-$ and $\phi$ candidates are shown in Fig.~\ref{fig:acc} (a,b). The number of signal counts is extracted using a bin counting method, and is shown as a function of $p_{\rm T}-y_{\rm cm}$ in Fig.~\ref{fig:acc} (c,d). Good mid-rapidity coverage is attained in the $\sqrt{s_{NN}}$ = 3 GeV Au+Au collisions.

\begin{figure}[h!]
\centering
\includegraphics[width=0.45\textwidth]{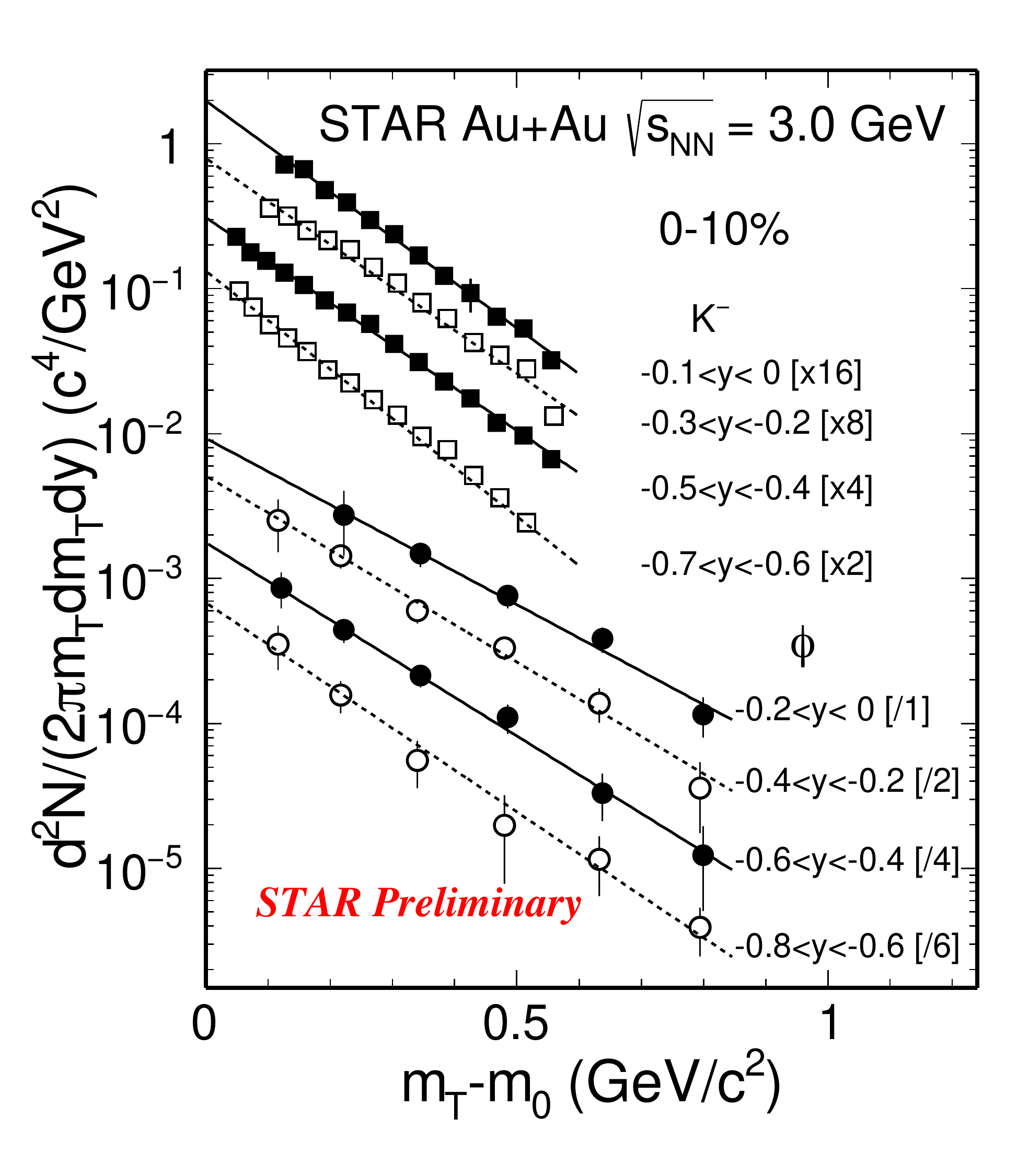}
\caption{$m_{\rm T}$ spectra for $K^-$ and $\phi$ in 0-10\% Au+Au collisions at $\sqrt{s_{NN}}$ = 3 GeV in different rapidity selections. The solid and dashed black lines represent fits using the $m_{\rm T}$ exponential function to the data points.}
\label{fig:mt}
\end{figure}

The numbers of $K^-$, $\phi$ and $\Xi^-$ counts in the data are extracted as a function of $m_{\rm T}-m_{0}$ in different rapidity and centrality selections. The raw signal counts in each $m_{\rm T}-m_{0}$ interval are subsequently corrected by the acceptance and reconstruction efficiency estimated via GEANT3 \cite{brun1987geant}. The corrected $K^-$, $\phi$, and $\Xi^-$ dN/d$m_{\rm T}$ are all well described by $m_{\rm T}$ exponential functions. The $m_{\rm T}$ and rapidity differential yield of $K^-$ and $\phi$ meson in 0-10\% centrality collisions are shown in Fig.~\ref{fig:mt}.

We considered three sources of systematic uncertainties, arising from (1) imperfect description of topological variables in the simulations, (2) the tracking efficiency of the TPC, and (3) the background subtraction method. Their contributions are estimated by varying topological cuts used in the analysis, the TPC track quality selection criteria and the background subtraction method. These uncertainties are assumed to be uncorrelated and added in quadrature. 

\begin{figure}[h!]
\centering
\includegraphics[width=0.9\textwidth]{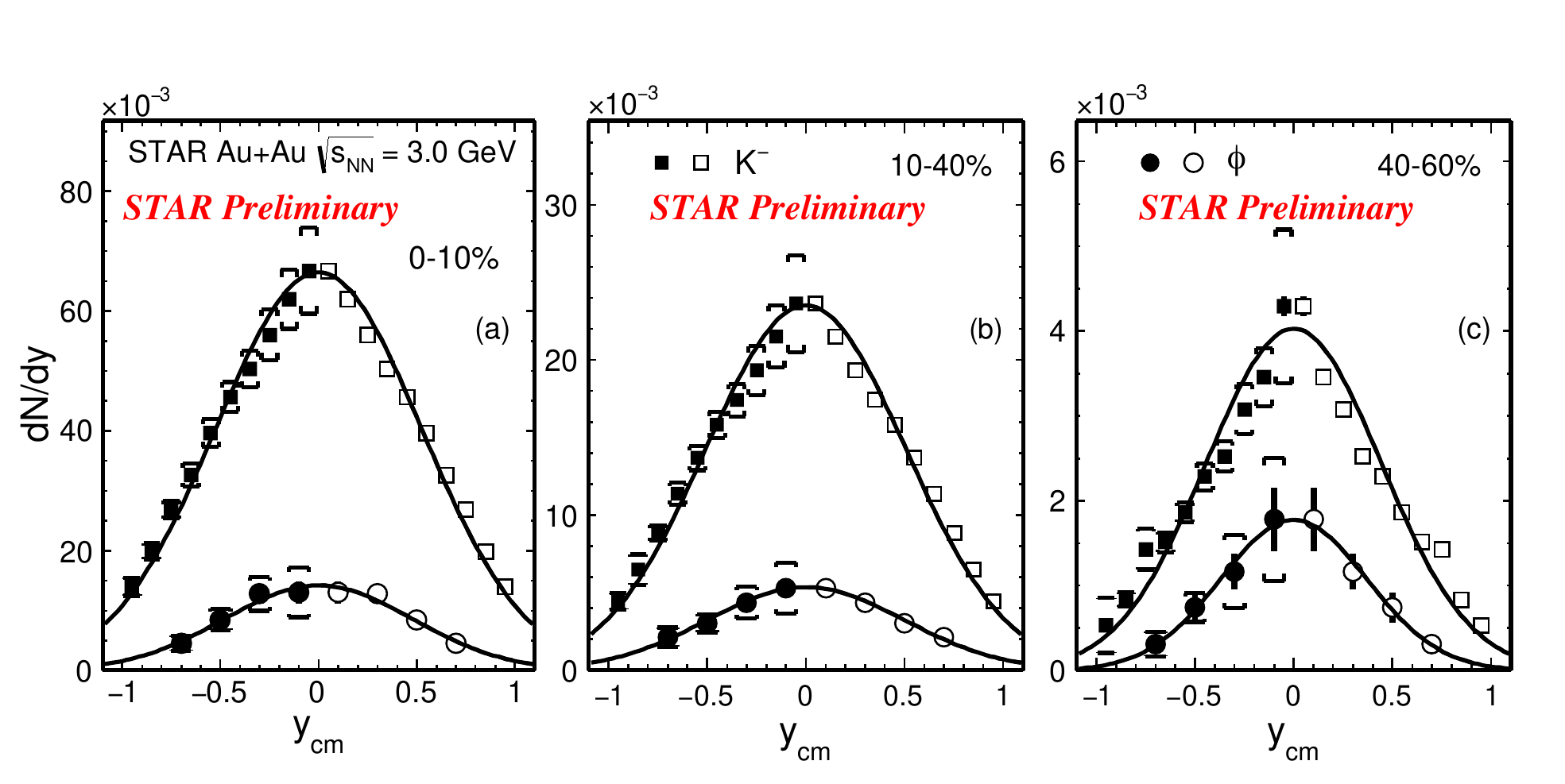}
\caption{Rapidity distributions of $K^-$ and $\phi$ meson for various centrality regions, solid symbols are measured data, open ones are reflection. Yields obtained from integrating fit functions to the $m_{\rm T}$ spectra are fit with a Gaussian function.}
\label{fig:dndy}
\end{figure}

To estimate the $p_{\rm T}$ integrated yield, the data are extrapolated down to $p_{\rm T}$ = 0. Besides the aforementioned systematic uncertainties, uncertainties from extrapolation to the unmeasured regions are considered, i.e. different functional forms are used for the extrapolation to estimate this uncertainty. The $p_{\rm T}$ integrated $dN/dy$ as a function of rapidity are shown in Fig.~\ref{fig:dndy}. Solid curves depict Gaussian function fits to the data points with the centroid parameter fixed to zero. They are used to extrapolate to the unmeasured rapidity region for calculating total multiplicities.

\section{Yield ratios}

\begin{figure}[h!]
\centering
\includegraphics[width=0.47\textwidth]{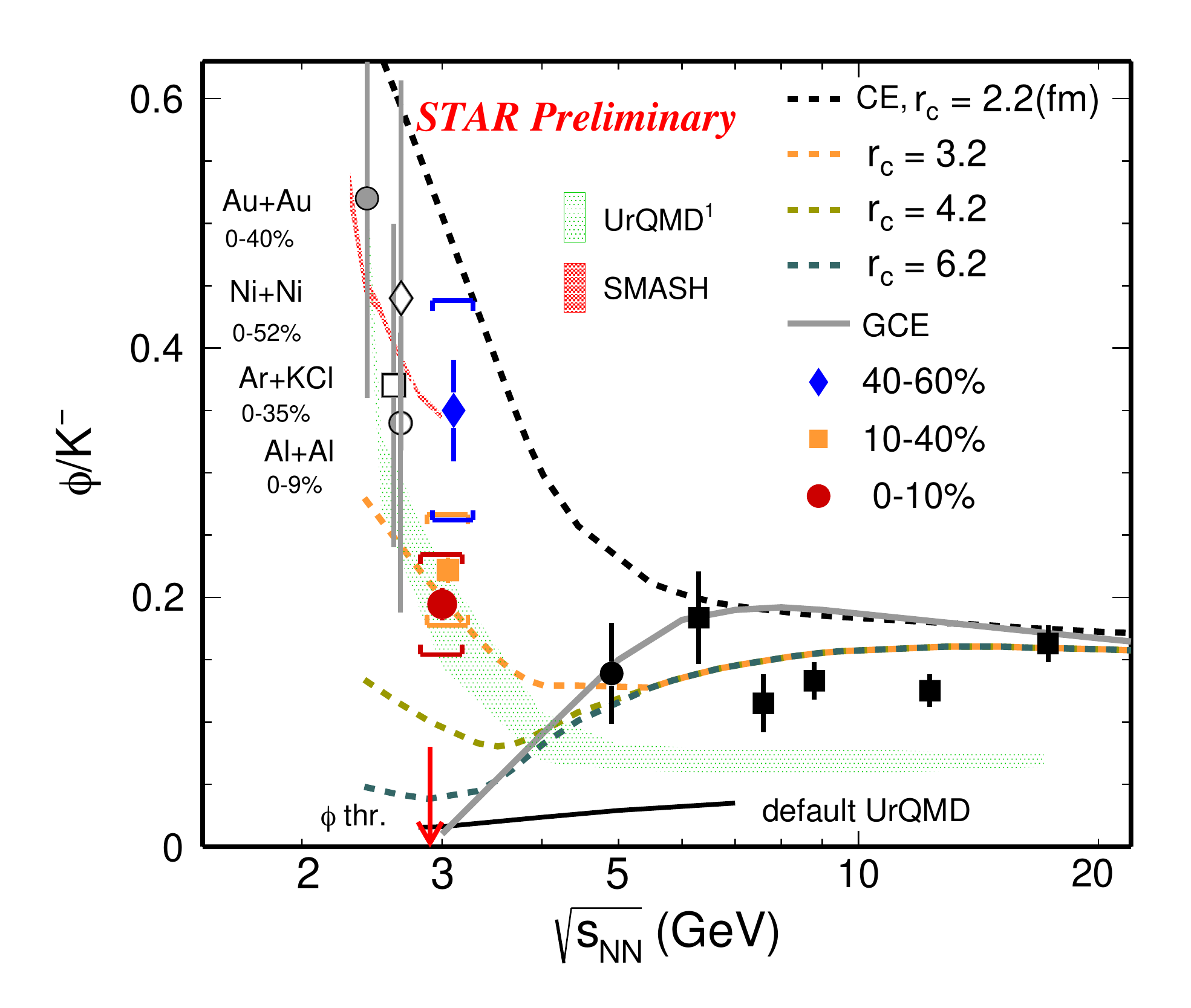}
\includegraphics[width=0.47\textwidth]{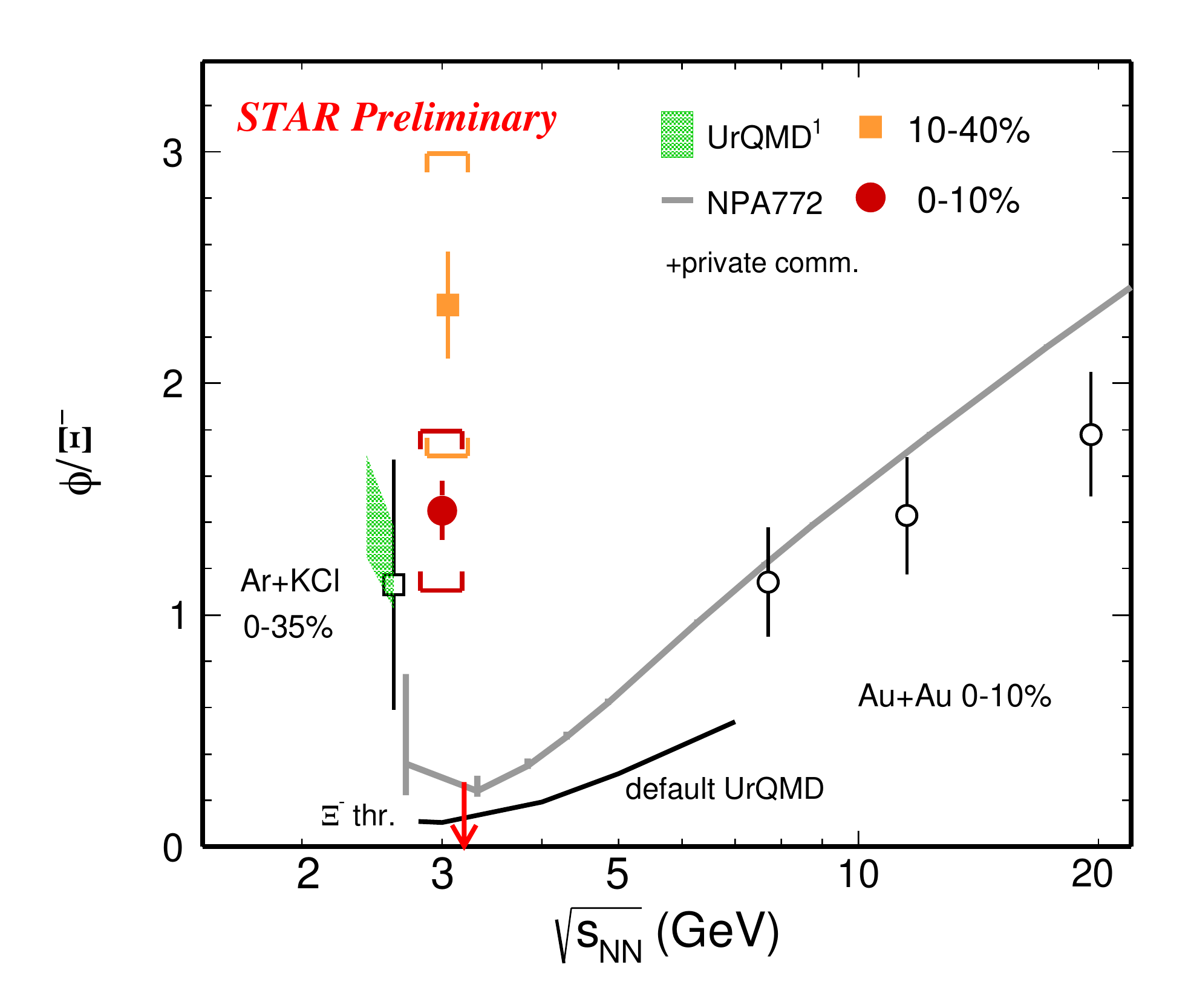}
\caption{$\phi/K^-$ (left) and $\phi/\Xi^-$ (right) ratios as a function of $\sqrt{s_{NN}}$. The solid color markers show the measurements at 3 GeV from different centrality bins with vertical lines for statistical and bracket symbols for systematic errors, while other markers are used for data from various other energies and/or collision systems \cite{back2004production, NA49:2008goy, STAR:2019bjj, gasik2016strange, piasecki2015influence, HADES:2009lnd, HADES:2017jgz, HADES:2009mtu}.}
\label{fig:ratio}
\end{figure}

Figure \ref{fig:ratio} shows the $\phi/K^-$ (left) and $\phi/\Xi^-$ (right) ratios from our measurements in different centralities as a function of $\sqrt{s_{NN}}$. The measured $\phi/K^-$ and $\phi/\Xi^-$ ratios at 3 GeV are slightly higher than the values at high energies for $\sqrt{s_{NN}}\geqslant$ 5 GeV. There is a hint that both the measured $\phi/K^-$ and $\phi/\Xi^-$ ratios in mid-central collisions are larger than those in central collisions, and this needs further statistics to systematically study the centrality dependence in detail. The GCE underestimates our data with $\sim 5\sigma$ effect for $\phi/K^-$ and $\sim 4\sigma$ effect for $\phi/\Xi^-$, while the CE calculation with strangeness correlation length ($r_{c}$) $\sim 3.2$ fm can reasonably describe our measurements. The precise determination of the thermal parameters (including $T_{ch}, ~\mu_{B}$ and $r_{c}$) needs a global thermal model fit with all the particle yields at 3 GeV. The modified transport models (UrQMD and SMASH) including high mass resonances decay to $\phi$ and $\Xi^-$ can also reasonably describe the data at low energies \cite{back2004production, NA49:2008goy, STAR:2019bjj, gasik2016strange, piasecki2015influence, HADES:2009lnd, HADES:2017jgz, HADES:2009mtu}. In the Au+Au collisions at 3 GeV, the observed strangeness production mechanism may be different from that at high energies, and this may indicate a change of Equation of State (EoS) at this low energy.

\section{Conclusion}

We report the systematic measurements of $K^-$, $\phi$ meson and $\Xi^-$ hyperon production yields and the $\phi$/$K^-$, $\phi/\Xi^-$ ratios in Au+Au collisions at $\sqrt{s_{NN}}$ = 3 GeV with the STAR experiment at RHIC. The measured $\phi/K^-$ ratio is significantly larger than the statistical model prediction based on GCE in the 0–10\% central collisions. Both the results of $\phi/K^-$ and $\phi/\Xi^-$ ratios favor the  Canonical Ensemble model for strangeness production in such collisions. Transport models, including the resonance decays, could reasonably describe our measured $\phi/K^-$ ratio at 3 GeV and the increasing trend of $\phi/\Xi^-$ at lower energies. These results suggest a significant change in the strangeness production at $\sqrt{s_{NN}}$ = 3 GeV compared  to higher collision energies, providing new insights towards the understanding of the QCD medium properties at high baryon density \cite{refId0}.

\section*{Acknowledgements}
This work was supported in part by the National Key Research and Development Program of
China under Grant No. 2020YFE0202002, and the Fundamental Research Funds for the Central
Universities under Grant No. CCNU20TS005.

\bibliography{SciPost_Example_BiBTeX_File.bib}

\nolinenumbers

\end{document}